# Enhanced optical trapping assisted by resonant energy backflow in a perforated dielectric microsphere

YURY E. GEINTS,[1,*] IGOR V. MININ,[2] OLEG V. MININ[2]

[1]*V.E. Zuev Institute of Atmospheric Optics, 1 Acad. Zuev square, Tomsk, 634021, Russia*
[2]*Tomsk Polytechnic University, Lenina 36, Tomsk, 634050, Russia*
*\*Corresponding author: ygeints@iao.ru*

## Abstract

Optical energy flow inside a dielectric microsphere exposed to an optical wave is usually codirected with its wave vector. At the same time, if the optical field in a microparticle is in resonance with a high-quality spatial eigenmode, referred to as the whispering-gallery mode (WGM), at least two regions of reverse energy flow emerge in the illuminated and shadow particle hemispheres. These areas are of considerable practical interest due to their enhanced optical trapping potential provided they should be previously cleared from particle material. In this paper, we consider a perforated microsphere with an air-filled pinhole fabricated along the particle diameter and theoretically analyze the conditions for WGMs excitation. A pinhole isolates the energy backflow regions of WGM and multiple enhances the optical pull-in force that transforms a perforated microsphere into an efficient optical tweezer for trapping various nanoobjects.

## 1. Introduction

Dielectric mesoscale particles are a family of wavelength-scaled optics possessing an elegant and predominantly subwavelength focusing of optical radiation with low diffraction losses [1-3]. Compared to metallic nanostructures, dielectric particles have a simple internal structure and are made from widespread available materials. They usually have weak intrinsic light absorption and allow for subwavelength radiation focusing without the need for expensive nanofabrication tools or complex near-field setups [2].

Subwavelength localization of the optical field incident on a particle leads to enhanced interaction of light with the environment, that is of great importance for many new industrial and scientific applications including surface nanopatterning [1], precise bio-sampling [4], cell biophysics [5], high-sensitive Raman spectroscopy [6, 7], nanolithography [8], etc. Undoubtedly, the ability to localize and concentrate light predominantly at the subdiffraction scale is of interdisciplinary nature and still attracts considerable scientific interest.

The specific nanostructuring the shadowed surface of dielectric microparticles by cutting out the nano-relief of various spatial shapes can provide for an improved localized field configuration for generating spatially-structured light beams for various practical purposes, among them, the nanoimaging, surface nanopatterning, light field generation with the optical angular momentum (OAM), producing the optical vortices, optical traps, etc. [9, 10-12]. Even a simplest type of a mesoscale particle nanostructuring by drilling a tiny nanohole (either dead-ended or open-ended) allows the field to be localized near the hole opening with the transverse size depending only on the size of this hole and the refractive index contrast of materials in the hole and particle regardless the optical wavelength of incoming wave (spherical particle with the refractive index $n = 2$ [13]). Thus, this approach makes it possible to form near-field spatially-localized beams with minimal waist beyond the diffraction limit, which opens up new prospects, in particular, in the field of optical manipulations with nanoparticles [14]. However, worthwhile noting that the all previously known works deal with the nonresonant mode of particle light scattering, which causes the so-called photonic nanojet formation near the particles with $n \leq 2$ [1-3, 9, 10-13].

The internal optical field in symmetric dielectric particles (sphere, cylinder, disc) exhibits resonances at precise frequency tuning of incident light wave to the frequencies of particle electromagnetic eigenmodes, the so-called, whispering gallery modes (WGMs). In this case the spatial structure of internal optical field is deeply restructured leading to sharp intensity increase caused by field concentration near the surface of a particle with formation of annular periodic structures resembling the standing wave pattern. In other words, a WGM is an optical resonance that is excited in an open optical microcavity possessing high spatial field localization close to the outer edge of the cavity and having extremely high-quality factor and high fidelity [15]. One of the interesting features which can be revealed in the regions of optical field concentration at resonant conditions is the appearance here of many optical vortices at the points of wave phase singularity, where a local reversal of the optical flux of energy directed toward the propagation of the incident wave can arise [16, 17]. A similar reversal of the energy flux is also observed during the sharp focusing of vortex singular beams by ordinary or gradient-index lenses [18, 19].

From the practical point of view, the phenomenon of reverse energy flow is of considerable interest because the optical beams exploited this effect can attract the nanobjects (including bacterial cells [20, 21]) towards the light source, forming the so-called, photonic "tractor beams" [22]. For example, in [23] the possibility of a reverse optical energy flow is demonstrated when the radiation goes through subwavelength slits fabricated in a metal plate. Such reversed optical currents appeared around the phase singularities formed near hole boundaries. Meanwhile, the inverse energy flux is also observed in perfect (non-structured) dielectric spherical particles when they are irradiated with a complex structured optical radiation, for example, a circularly polarized Gauss-Laguerre beam [24]. However, the phenomenon of reverse energy flux in the region of subwavelength field localization near the surface of mesoscale particles, especially under WGM resonances, is still rarely studied.

In this paper, following our recent work [17] we consider the optical energy backflow appearance in a dielectric microsphere being nanostructured by an open-ended hole fabricated along the particle principal diameter. A distinctive feature of the case considered is that the nanostructured microparticle experiences the excitation of internal field resonances. By means of the finite element method implemented in the COMSOL Multiphysics solver, the numerical simulations of the near-field optical wave diffraction in the resonant mode are carried out. We present the detailed analysis of the spatial distribution of the Poynting vector in the "hot areas" located in the shadow and illuminated hemispheres of the microparticle. It turns out that the reversal of the optical energy flux occurs inside the hole near its open-end, and the backflow amplitude is several times greater than the value of the direct energy flux and depends on the width of the nano hole and the quality factor of the WGM being excited. This circumstance allows us proposing the concept of an efficient optical trap based on a perforated dielectric microsphere, where the trapping region is formed inside the hole (inside the particle). Importantly, the optical "tractor" (pull-in) forces acting on a test nanoparticle are significantly enhanced by the resonant reverse energy flux.

## 2. Numerical model of a perforated sphere

Consider the following setup of the problem solved. As shown in Fig. 1(a), a spherical dielectric microparticle with the radius $R$ and refractive index $n$ is located in air and is illuminated by a plane monochromatic circularly polarized electromagnetic wave propagating in the direction of the wave vector $\mathbf{k}_z$. Note that the choice of such wave polarization state is only due to the axisymmetric formulation of the electromagnetic problem and does not lead to any losses of generality of the results obtained. Inside the microsphere, along its main diameter (in the direction of light incidence) a cylindrical hole with the diameter $d_h$ is fabricated (see, Fig. 1(b)) and is filled with air with the refractive index $n_1 = 1$. All simulations are performed for a dielectric microsphere with $n = 2$ and radius $R = 1$ µm, which in practice can be made, e.g., from polycrystalline barium borate (OHARA L-BBH1 [25]). The incoming optical wavelength $\lambda$ is chosen in the visible spectral range where no optical absorption of the particulate material is present. The dimensionless diffraction parameter $\rho = 2\pi R/\lambda$, known as the particle Mie size parameter, in this case equals to about $4\pi$, which the particle considered to the mesowavelength range [2].

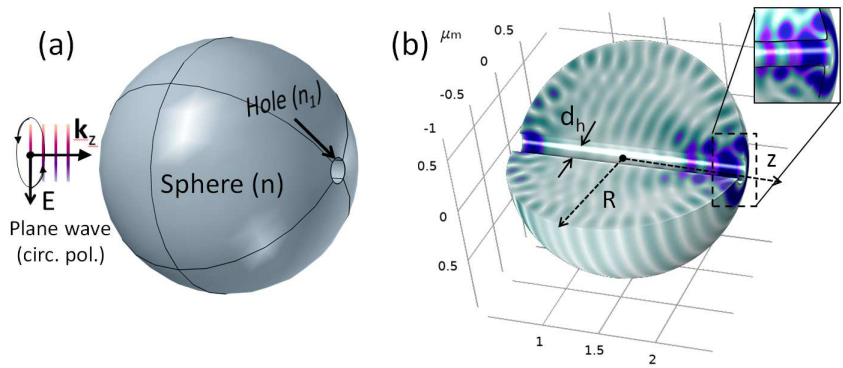

Fig. 1. (a) Schematic of a numerical model of light scattering on a perforated microsphere (microbead); (b) 3D-view of the optical field structure |**E**| inside the bead. A close-up of the rounded area near the hole exit is shown in the inset.

The spatial distribution of the optical field inside and around the nanostructured microparticle is calculated in the stationary approximation through the numerical solution of the Helmholtz equation using the finite element method implemented in the COMSOL Multiphysics package. To avoid the undesirable extreme field localization near the sharp edges of particle nanorelief [13], the outlet region of the nanohole on the surface of the microsphere is preliminary rendered using the COMSOL geometric fillet operation which produces the rounded edges as shown in the close-up image in the inset to Fig. 1(b). The spatial resolution of the tetrahedral mesh is chosen less than $\lambda/(15n)$ in the region outside the hole and $d_h/(5n_1)$ inside it. All boundaries between heterogeneous domains are meshed in multiples of the smallest grid step to improve the accuracy of the solution. At the outer boundaries of the computation domain, the conditions of wave free propagation are set by constructing a system of perfectly matched layers (PML) minimizing the parasitic wave reflections.

## 3. Poynting vector and optical backflow simulations

First, we analyze the physical reasons for the optical energy flux reversal in the nanohole region at wave diffraction on a perforated sphere. To this end, consider the spatial distribution of energy flows using the time-averaged Poynting vector field $\mathbf{S} = (c/8\pi)\,\mathrm{Re}\left[\mathbf{E}\times\mathbf{H}^*\right]$, where $\mathbf{E}$ and $\mathbf{H}$ are electric and magnetic field vectors, respectively, $c$ is the speed of light. In Figs. 2(a)-(c), the 2D spatial distribution of the electric field norm $|\mathbf{E}|$ and Poynting vector $\mathbf{S}$ in the vicinity of the microbead are shown when illuminated by a plane optical wave with $\lambda = 532$ nm. The cylindrical coordinates $(r,\phi,z)$ are used.

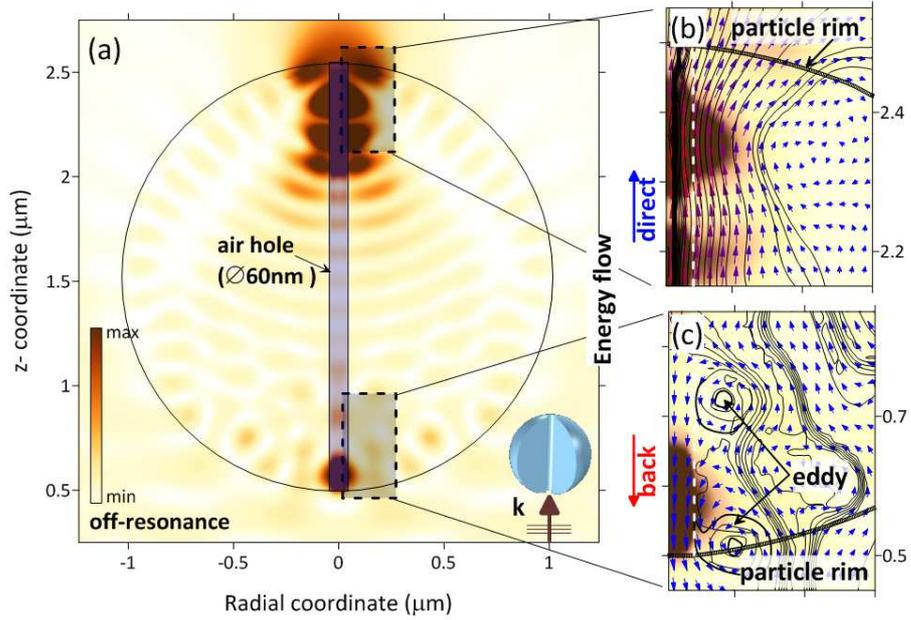

Fig. 2. Spatial distribution of (a) electric field norm $|\mathbf{E}|$ (color maps) inside a microbead and (b) Poynting vector $\mathbf{S}$ (contours and arrows) in the shadow (b) and illuminated (c) nanohole end regions. In (b) and (c), the hole boundary is shown by the white dashed line.

As follows from this figure, both the direction and the very nature of the energy flows can change in the regions of the exit end of the hole. Since the focusing region is located near the interface between the particle and the environment, it can be used to separate the areas of forward and backward energy fluxes [26]. According to the physical nature, the Poynting vector (optical energy) flux can be subdivided to a potential (orbital) $\mathbf{S}_O$ and a solenoidal (spin) $\mathbf{S}_C$ fluxes [27, 28]: $\mathbf{S} = \mathbf{S}_O + \mathbf{S}_C$. The orbital Poynting component, which is directly related to the wave phase gradient $\partial\varphi/\partial z$, is responsible for the appearance of the inversely directed energy current $S_z \equiv \mathbf{e}_z \cdot \mathbf{S}_O = I/k\,(\partial\varphi/\partial z) < 0$, where $I = c/8\pi|\mathbf{E}|^2$ is the field intensity. Thus, in the regions with negative longitudinal component of the phase gradient $\partial\varphi/\partial z$ one should expect a reverse of the energy flow. In the case considered, this is the region of field focusing near the illuminated hemisphere of the particle (see, Fig. 2c), because here the optical field singularities in the form of vortices are evident with the optical energy moving in a counterclockwise direction. At the same time, in the shadow hemisphere of the microparticle (Fig. 2b) the energy flow inside the air hole is laminar and is co-directed with the incident wave vector, and no vortex energy motions are formed in this zone [13].

The situation considered above referred to the case of the nonresonant optical field scattering on a dielectric spherical particle. Now turn to the case with resonant excitation of a WGM in the particle. This is illustrated in Fig. 3(a,b) for WGM with the $TE_{19,1}$ polarization having the resonance wavelength $\lambda$ = 530.04 nm.

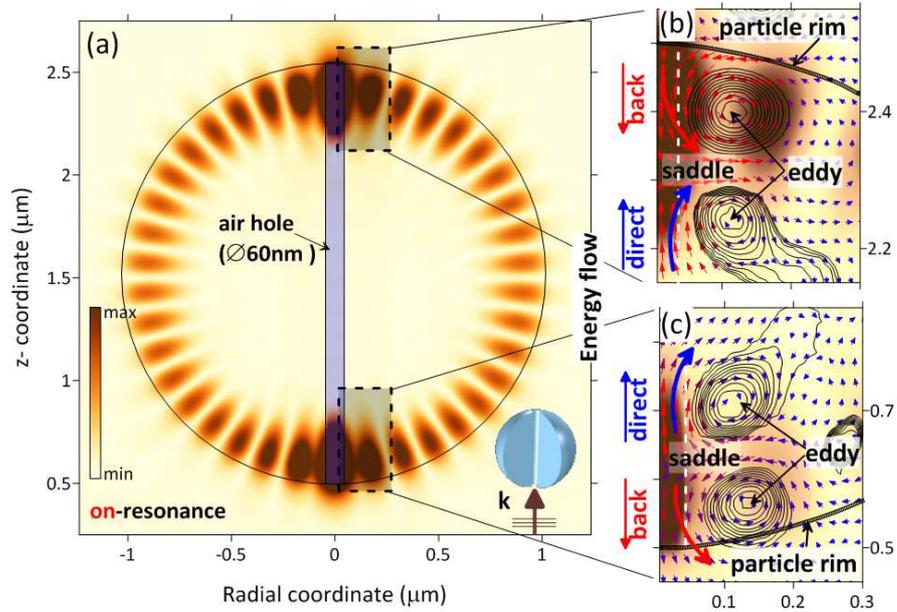

Fig. 3. Same as in Fig. 2 for $TE_{19,1}$ WGM resonant excitation.

As seen, under the conditions of the eigen resonance the distribution of the optical field intensity experiences deep spatial transformations by taking the characteristic shape of a standing wave localized along the sphere rim. Besides, the structure of the energy flows near the hole ends is changed also. Here, the pronounced paired singular regions arise with counter-directional vortex motion of optical energy separated by a region with radial energy flow forming a special saddle-type phase singularity [27]. In this case, in the vicinity of such singular points which have the essential subwavelength dimensions (see the inserts to Fig. 3 showing the size of singular points of about 0.1 µm or $\lambda/5$) the values of the local wave vector can reach huge values due to the sharp wave phase gradient [29].

Due to the mirror-opposite direction of optical vortices rotation in the shadow (Fig. 3b) and illuminated parts of the microsphere (Fig. 3c), the optical energy flows into the annular region of the eigenmode through the "shadow" saddle. On the contrary, through the saddle in the illuminated part of the particle, the WGM energy enters the interior of the air hole forming the divergent forward and reverse currents of the Poynting vector (indicated in the figures by blue and red bold arrows). We can say that the paired optical vortices act like the gears of a conveyor belt or caterpillar track providing a circular energy flow inside the spherical particle eigenmode. Such spatial arrangement of optical vortices is closely related to the so-called optical nanomotors [30, 31] which in principle can be used for environment material transport and mixing.

In the context of optical manipulation of various nanoobjects, the area in the shadow part of the microsphere is of greatest interest since the presence here of a reversed energy flow provides an increase in the retraction mechanical force of the optical tweezers due to the sign reversal of light pressure which is proportional to the value of $S_z$ [32]. The direct comparison of the amplitudes of the longitudinal Poynting component $S_z$ arising at the hole outlet of holes with different diameters $d_h$ for the several WGMs in the microsphere is shown in Fig. 4(a). For reference, we also plot the $S_z(z)$ profile for a solid microparticle (without a hole) at $TE_{19,1}$-mode resonance (gray dashed curve) as well as for a particle with a 60 nm (~0.1$\lambda$) hole but not in resonance (as in Fig. 2).

It is worthwhile noting that structuring a hole in a solid dielectric particle changes the WGM resonance conditions by shifting the resonance wavelength to the blue spectral region. Qualitatively, this result can be explained by a decrease in the effective refractive index of the hole particle, $n_{eff} \approx n(1 - V_h/V_0)$, where $V_0$, $V_h$ are the volumes of sphere and hole, respectively. Since the resonance Mie-parameter $\rho_0 = 2\pi R/\lambda_0$ of particle is inversely proportional to the refractive index [33] following the known dependence $\rho_0 \approx (l^2 - 1)/n$, where $l$ is the eigenmode number (in this case $l = 19$), then the resonance wavelength $\lambda_0$ will be also proportional to $n$ ($\lambda_0 \propto n$). Thus, considering the decrease in the effective refractive index of the perforated particle, we obtain that the change $\delta\lambda_0$ in the resonance WGM position is proportional to the relative volume of the hole: $\delta\lambda_0 \propto -V_h/V_0 \propto -(d_h/R)^2$. This parabolic dependence can be clearly seen in Fig. 5(b).

Note a similar trend in the scattering characteristics of perforated microspheres reported earlier in [34] for particles with a refractive index near unity ($n = 1.05$). Besides, in [35] a significant shift of WGM spectral contour in a solid dielectric microparticle is revealed, when a small metallic nanoparticle is placed near master particle surface because of the strong field coupling between WGM and plasmonic resonance. The spectral shift and splitting of dielectric microsphere eigenmodes in the presence of a Rayleigh-sized defect are discussed in detail in [36].

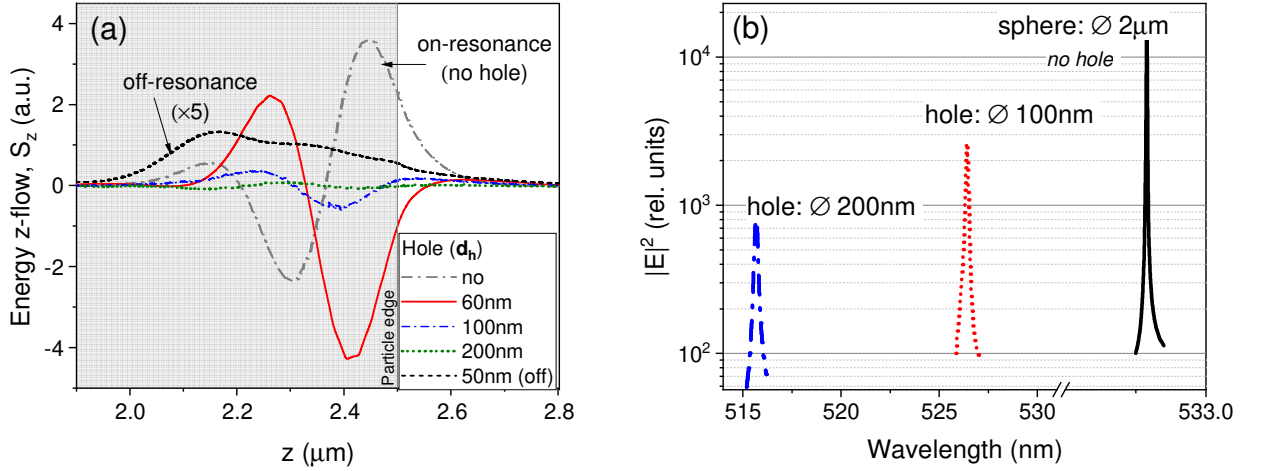

Fig. 4. (a) Longitudinal Poynting component $S_z$ inside a hole as calculated in the shadow microsphere region in the dependence on hole diameter $d_h$. The coordinate region $z < 2.5\mu m$ corresponds to sphere interior (shaded). (b) Dynamics of $TE_{19,1}$ spectral contour with hole diameter variation.

Recalling Fig. 4(a), we note that when the incident radiation is not in resonance with the sphere WGM (dashed line, values are magnified by factor of 5), no reverse energy current is organized in the hole area. The longitudinal component of the Poynting vector is permanently positive throughout the air hole, $S_z > 0$. However, at resonance, a small and initially positive $S_z$ flux in air abruptly changes its sign and becomes oppositely directed to the incident wave even before the exit slice of the hole ($z = 2.5\mu m$). Noteworthily, with a hole diameter of 60 nm, the ratio of the peak values of $|S_z|$ in the regions of negative (at $z = 2.4$ μm) and positive (at $z = 2.26$ μm) optical current is about two times. This is close in magnitude to the corresponding ratio of inverse and forward Poynting fluxes in the situation with the focusing of a second-order cylindrical vector beam by a Mikaelian gradient lens with a fabricated air crater [19] when the lens size was two times larger than the considered mesoscale particle Mie-size parameter $q \sim 38$. However, the optical scheme considered in our study is significantly simpler for the implementation and does not require specially structured optical beams and a specific complex structure of the refractive index of a scatterer.

The peak $|S_z|$ values in the reverse energy flux region when the hole diameter changes are summarized in Fig. 5(a). The inset to this figure shows the dependence of the maximum optical intensity on hole diameter $d_h$.

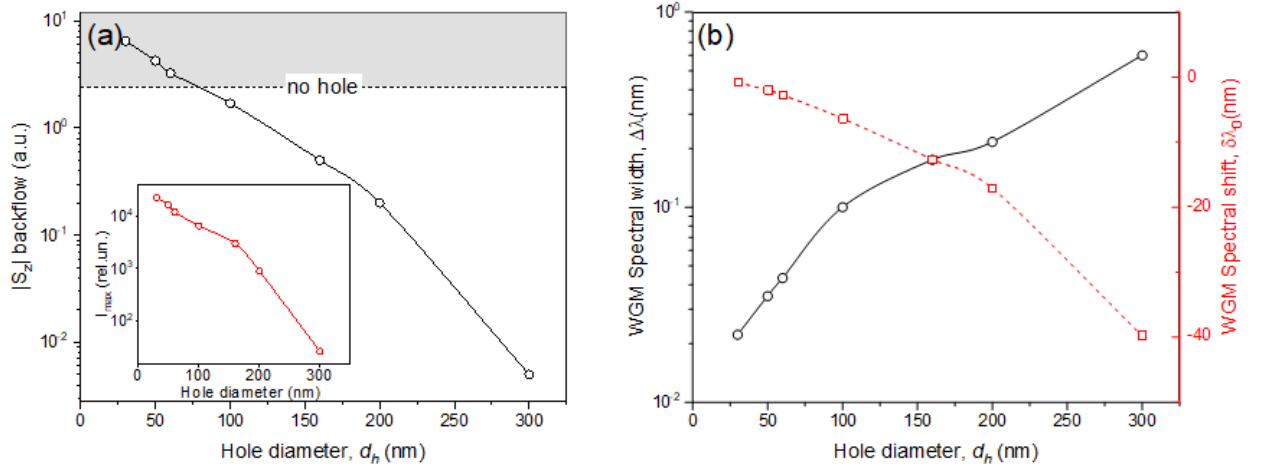

Fig. 5. (a) Longitudinal Poynting flux amplitude $|S_z|$ in the area of reversal energy flow; (b) Spectral width $\Delta\lambda$ and shift $\delta\lambda_0$ of TE$_{19,1}$- eigenmode depending on the hole diameter $d_h$. Inset: Peak intensity $I_{max}$ inside the hole in shadow part of the particle.

As seen, the amplitude of the Poynting backflow decreases sharply with increasing the hole diameter. The regression of this dependence gives an descending exponential dependence, $|S_z| \propto \exp\{-ad_h\}$, with a decrement $a \approx 21$ pm that correlates with a decrease in the WGM intensity in the hole area according to the corresponding law: $I_{max} \propto \exp\{-bd_h\}$, with a somewhat smaller but close in magnitude decrement, $b \approx 18.7$ pm. This is clear, because by virtue of its definition the longitudinal Poynting component is directly proportional to the optical field intensity, $|S_z| \propto I$, and the intensity drop, in turn, is associated with a broadening of the spectral contour $\Delta\lambda$ of mode resonance (Fig. 5(b)) and its quality factor decreasing as hole diameter $d_h$ increases. That is why, tuning to another, lower quality eigenmode of the microparticle leads to a decrease of the Poynting backflow in the area of the hole outlet. Indeed, our simulations show that by exploiting the TE$_{18,3}$-mode excited in a microsphere with a 60-nm aperture hole, the amplitude of the opposite energy flux $|S_z|$ becomes almost ten times smaller than in the case shown in Fig. 5(a). At the same time, for a 100 nm (~$\lambda/5$) hole this ratio is already more than 20 times. Thus, the highest values of the energy backflow $|S_z|$ in a hole are realized for the highest-order WGMs.

## 4. Optical trapping dynamics in the energy backflow region of a perforated sphere

In this section, we analyze the optical trapping forces acting on a test nanoparticle placed near a dielectric hole-structured micron-sized sphere in the region of hole opening. A gold nanoball with diameter $D_b$ = 30 nm and refractive index $n_b$ = 0.56-$i$·2.2 (at the wavelength 530 nm [37]) is used as such a test object. The mechanical optical forces acting on a solid object are determined through the full-wave calculations of the Maxwell stress tensor **T** which is related to the Poynting vector flow **S**. In an isotropic medium, the following expression for the integral optical force are used provided that the optical fields are temporally averaged [38]:

$$\mathbf{F} = -\oiint_S (\mathbf{T} \cdot \mathbf{e}_n) d\sigma \quad (1)$$

$$T_{ij} = \frac{1}{2}\delta_{ij}\left(\varepsilon_0|\mathbf{E}|^2 + |\mathbf{H}|^2\right) - \left(\varepsilon_0 E_i E_j + H_i H_j\right) \quad (2)$$

Here the integral in Eq. (1) is taken along the surface of a test particle with the outward normal $\mathbf{e}_n$, whereas $T_{ij}$ are the components of the stress tensor. The medium and the particles are assumed to be non-magnetic.

In paraxial approximation, when it is possible to distinguish the preferred direction of optical wave propagation ($\mathbf{k}_z$), the expressions for optical force (1)-(2) formally may be represented as a sum of two forces [39]: $\mathbf{F} = \mathbf{F}_s + \mathbf{F}_g$. Here, $\mathbf{F}_s$ is the dissipative optical force, or so-called "scattering force", which is due to the pressure of optical radiation on particle surface and is associated with the angular field momentum. $\mathbf{F}_g$ is the gradient optical force associated with the Lorentz force acting on the charge in the electric field which is caused by the non-uniform optical intensity distribution in space. The first optical force acts in the direction of the orbital energy flux $\mathbf{S}_O$ [32], $\mathbf{F}_s \sim \mathbf{S}_O$, while the gradient force is collinear with direction of the electric field gradient averaged over the nanoparticle surface, $\mathbf{F}_g \sim \nabla|\mathbf{E}|^2$. Meanwhile, under normal conditions the orbital energy flux is co-directed with the propagation of the optical wave, i.e., the $\mathbf{k}_z$ vector, and the scattering force always tends to push the trapped particle out of the gradient optical trap. However, in the regions where the energy flow is reversed, $\mathbf{S}_{O,z} \downarrow\uparrow \mathbf{k}_z$, the dissipative force acts towards the wave propagation and tends to pull the test object into the hole [14]. Obviously, this will increase the potential for optical trapping of different-kind nanoobjects in various practical applications, e.g., in the optical surface cleaning [26].

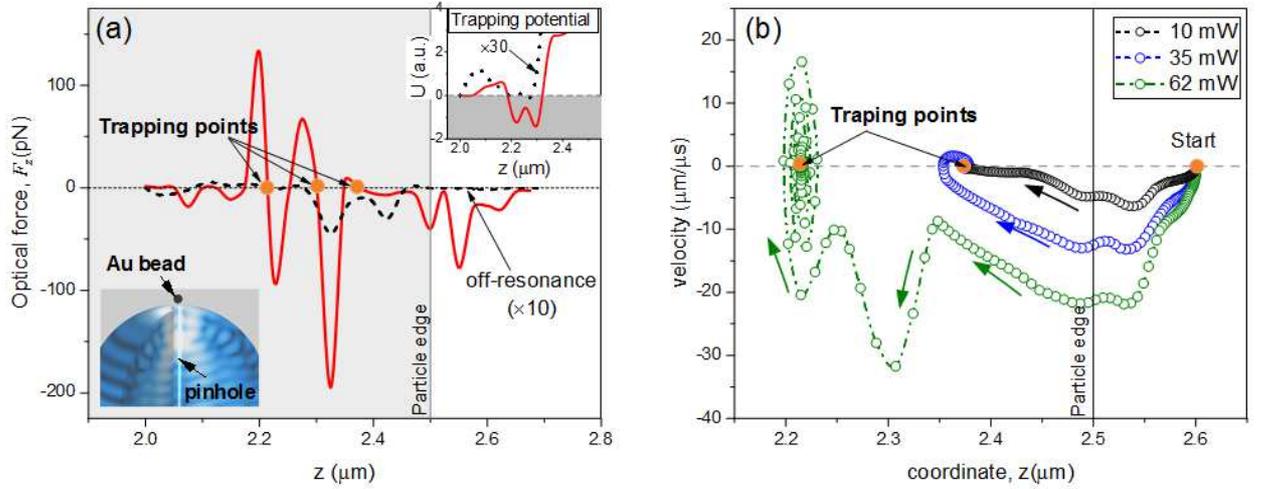

Fig. 6. (a) Longitudinal optical force $F_z$ acting on a 30 nm Au-bead placed near a dielectric microsphere structured by a 60 nm hole at WGM excitation in on-resonance (red solid line) and off-resonance conditions (black dashed line). Inset: Longitudinal optical trapping potential $U$. (b) Phase diagrams of Au-nanoball motion in proposed resonant optical trap based on a perforated microsphere upon laser illumination with different initial power.

Fig. 6(a) shows the profiles of net longitudinal optical force $F_z$ acting on a gold nanosphere located in the trapping region formed near the shadow part of a dielectric microsphere ($n = 2$) with a 60 nm through hole. In the calculations, a light beam with the power 10 mW is focused into a waist with a cross-section of 10 µm$^2$.

Importantly, when TE$_{19,1}$ WGM is excited in a microparticle (red solid curve), the optical force acting on a nanobead becomes negative approximately at the distance $\lambda/2$ from the surface of dielectric particle ($z = 2.5$ µm), i.e., the gold nanoball will be pulled inside the hole. The peak value of the pulling force in this case reaches ~ 200 pN, which is an order of magnitude higher than a similar value in an optical trap constructed on the photonic nanojet (PNJ) effect [40] and two orders of magnitude higher than in the optical scheme with beam focusing by a hole-structured gradient lens [41] at close values of optical pulse intensity (~ 0.1 MW/cm$^2$). Inside the hole, $F_z$ exhibits periodically sign change and the optical force actually disappears beyond about one laser wavelength when going deeper inside the hole. The oscillatory $F_z$ behavior leads to the formation of three trapping points (shown by yellow circles), two of which are located at $z = 2.2$ µm and $z = 2.2$ µm, respectively, and coincide with the potential wells of the optical trapping function: $U(z) = -\int^{z} F_z dz'$, as shown in the inset to Fig. 6(a).

For comparison, the optical capture force and trapping potential in the case when the optical field inside the microparticle is nonresonant (black dashed lines) are also plotted in this figure. In this case, the nanobead can also be trapped in the region $z = 2.28\mu m$, but the optical pull-in force is approximately 25 times smaller than in the resonant WGM excitation. Worthwhile noting, *H. Wang et al.* [42] reported the specific optical force of 0.2 pN/W for a 20-nm particle trapped near a 3μm dielectric cylinder at a WGM resonance. However, this optical force reaches only about 2 pN if applied to a laser beam power of 10 mW which is two orders of magnitude lower than in the proposed here optical trap.

To illustrate the potential of an optical trap based on a perforated microsphere, we numerically solve the equation of motion of a small solid particle (nanobead) with mass $m$ in the Earth's gravitational field under the action of an external optical force. In accordance with the first Newton's law one can write down the following equation:

$$m\frac{d^2z}{dt^2} = F_z - \gamma\frac{dz}{dt} - mg \qquad (3)$$

Here $F_z$ is the optical force, $t$ is time, $\gamma = 3\pi\eta D_b$ is Stokes coefficient of friction for sample particle moving in a homogeneous medium with dynamic viscosity $\eta$ (in air, $\eta = 1.8 \cdot 10^{-5}$ kg/(m·s)), $mg$ is gravity projection acting in the negative $z$-direction (gravity constant, $g = 9.8$ m/s$^2$). The solution of this equation is carried out numerically in the axisymmetric 2D-domain. The optical force is recalculated each time the position of the particle changes. The electromagnetic field vectors are obtained using the FEM technique (COMSOL Multiphysics) with an adaptive time stepping. At the initial time moment, a gold sphere with a diameter of 30 nm is placed near the microsphere at a point with the coordinates ($r = 0$, $z = 2.6\mu m$) and then the nanobead begins to move under the action of the net mechanical force toward the hole region.

Fig. 6(b) shows the phase portrait of nanoparticle motion in the proposed optical trap at TE$_{19,1}$ - mode excitation in a microsphere exposed to a laser radiation with different power and $\lambda = 530.04$ nm. Recall that the phase portrait is a trajectory in the phase space showing the dependence of material object velocity on its spatial position (in this case, $z$ coordinate). As seen, depending on laser beam power, the nanobead starts moving from the starting point ("Start") located at $z = 2.6\mu m$ towards the hole opening along different phase trajectories, i.e. with different initial velocities. After a few microseconds, the nanoparticle reaches the first trapping point inside the hole, where it then relaxes near the end point of its trajectory. After a while, the movement of the gold particle at this point completely stops. Interestingly, due to the presence of several minima of the optical trapping potential profile $U(z)$, the test nanosphere can gain more inertia and skip the trapping points near the hole exit and relax in the deeper trapping zone providing sufficient irradiation power. This situation is illustrated by the phase trajectory for a laser power of 62 mW, when the nanosphere consistently passes the outer trapping points at $z = 2.38\mu m$ and $2.3\mu m$, and after a series of oscillations finally stops at third trapping point with $z = 2.2\mu m$ deep inside the hole.

## 5. Conclusions

To conclude, we have considered a specific optical diffractive element representing a wavelength-scale dielectric sphere with a through nano-hole structured along sphere diameter. Using the finite element method, we numerically simulate the near-field diffraction of a circularly polarized optical wave on perforated microparticle and study in detail the spatial structure of the optical field energy flow (the Poynting vector) in the region of the hole ends. It turned out that the perforated microsphere, as well as the solid particle, supports excitation of high-quality electromagnetic eigenmodes (WGMs). The presence of a hole causes the blue-shift of WGM resonant wavelength, as well as the quality factor and eigenmode intensity decrease. Importantly, at a WGM resonance, in the region of the hole, a strong flux of optical energy arises inside the particle, which is directed opposite to the wave incidence. The intensity of this resonant energy backflow decreases with increasing the hole diameter and decreasing the quality factor of the excited resonance, but remains multiples of the value achieved in the case of nonresonant light scattering.

The presence of regions with optical current reversal near a microsphere is promising for optical trapping of nanoscale objects, which can be pull inside the air hole by the field gradient and dissipative optical forces, thus transforming a perforated microsphere into an "optical vacuum cleaner" [14]. Under the conditions of internal field resonance, the efficiency of such an optical trapping increases dramatically. A numerical experiment with a 30 nm gold sphere shows that the specific retraction optical force in this case can reach 20 pN/W.

Currently, the manufacturing of a mesowavelength dielectric sphere with a cylindric hole is possible, e.g., using the material removal by the ablation process under the influence of photonic nanojet, when the heat is released in the focusing region located in the shadow hemisphere. In this case, the refractive index contrast of the particle material must be at least a critical value $n = 2$, so that the focusing region does not extend beyond the particle interior [2]. Due to the effect of heat accumulation in this region, the formation of a microcrater with direct removal of the particle material occurs [43, 44].

**Funding.** Ministry of Science and Higher Education of the Russian Federation (IAO SB RAS).

**Acknowledgments.** I.M. and O.M. thank Tomsk Polytechnic University Development Program for supporting this work.

**Disclosures.** The authors declare no conflicts of interest.

**Data availability**. Data underlying the results presented in this paper may be obtained from the authors upon reasonable request.

**Author contributions.** Y.Geints: Investigation, writing -original draft, software, visualization. I.Minin and O.Minin: Conceptualization, methodology, data curation, validation.